\documentclass[preprintnumbers,amsmath,amssymb,twocolumn]{revtex4}
\usepackage{bm}
\usepackage{graphicx}% Include figure files
\usepackage{dcolumn}

\begin{document}

\title{Quasiparticle Excitations in the Superconducting State of FeSe Probed by Thermal Hall Conductivity in the Vicinity of the BCS-BEC Crossover}

\author{T.\,Watashige$^{1}$}
\altaffiliation{These authors contributed equally to this work.}
\author{S.\,Arsenijevic$^{2,\ast}$}
\author{T.\,Yamashita$^1$}
\author{D.\,Terazawa$^1$}
\author{T.\,Onishi$^1$}
\author{L.\,Opherden$^2$}
\author{S.\,Kasahara$^1$}
\author{Y.\,Tokiwa$^1$}
\author{Y.\,Kasahara$^1$}
\author{T.\,Shibauchi$^3$}
\author{H.\,v. L\"{o}hneysen$^4$}
\author{J.\,Wosnitza$^2$}
\author{Y.\,Matsuda$^1$}

\affiliation{$^{1}$ Department of Physics, Kyoto University, Kyoto 606-8502 Japan}
\affiliation{$^{2}$ Hochfeld-Magnetlabor Dresden (HLD-EMFL), Helmholtz-Zentrum Dresden-Rossendorf, D-01314 Dresden, Germany }
\affiliation{$^3$ Department of Advanced Materials Science, University of Tokyo, Chiba 277-8561, Japan} 
\affiliation{$^4$ Physikalisches Institut, Karlsruhe Institut f\"{u}r Technologie, Wolfgang-Gaede-Str.\,1, D-76131 Karlsruhe, Germany} 
%\date{\today}% 

\begin{abstract}
There is growing evidence that the superconducting semimetal FeSe ($T_c\sim8$\,K) is in the crossover regime between weak-coupling Bardeen-Cooper-Schrieffer (BCS) and strong-coupling Bose-Einstein-condensate (BEC) limits.  
%Therefore FeSe offers  a unique and fascinating platform to study the crossover physics in the multiband  systems.  
%Here, to obtain information on the quasiparticle excitations in the superconducting state of FeSe, 
We report on longitudinal and transverse thermal conductivities, $\kappa_{xx}$ and $\kappa_{xy}$, respectively, in magnetic fields up to 20\,T. The field dependences of $\kappa_{xx}$ and $\kappa_{xy}$ imply that a highly anisotropic small superconducting gap forms at the electron Fermi-surface pocket whereas a more isotropic and larger gap forms at the hole pocket. Below $\sim1.0$\,K, both $\kappa_{xx}$ and $\kappa_{xy}$  exhibit distinct anomalies (kinks) at the upper critical field $H_{c2}$ and at a field $H^*$ slightly below $H_{c2}$. The analysis of the thermal Hall angle ($\kappa_{xy}/\kappa_{xx}$) indicates a change of the quasiparticle scattering rate at $H^*$.  These results provide strong support to the previous suggestion that above $H^*$ a distinct field-induced superconducting phase emerges with an unprecedented large spin imbalance.  % The strongly band-dependent quasiparticle structure and the emergent highly spin-polarized  phase may be important and unique features of the BCS-BEC crossover in multiband superconducting system. 
\end{abstract}

%%% Keywords are not needed any longer. %%%
%%%\kword{keyword1, keyword2, keyword3, \ldots}
%%%

\maketitle

\section{Introduction}

The physics of the crossover between the weak-coupling Bardeen-Cooper-Schrieffer (BCS) and strong-coupling Bose-Einstein-condensate (BEC) limits gives a unified framework of quantum bound (superfluid) states of interacting fermions  \cite{Nozi85,Legg80}.  In the region of this crossover, the size of the interacting pairs, which is represented by the coherence length $\xi$, becomes comparable to the average distance between particles, which is close to the inverse Fermi momentum $1/k_F$, i.\,e., $k_F\xi\sim 1$ \cite{SadeMelo2008,Chen05,Gior08,Rand14}.  The BCS-BEC crossover has hitherto been realized experimentally in ultracold atomic gases.  On the other hand, in solids, almost all superconductors are in the BCS regime, where the superconducting gap energy $\Delta$ is usually several orders of magnitude smaller than the Fermi energy $\varepsilon_F$, $\Delta/\varepsilon_F\sim 1/k_F\xi\ll1$, and it has thus been extremely difficult to access the crossover regime. 

The iron-chalcogenide compound FeSe, comprised of two-dimensional Fe-Se layers~\cite{Hsu08}, has recently stirred great interest because it exhibits several unique properties.  At ambient pressure, the structural transition from tetragonal to orthorhombic crystal symmetry (nematic transition)  occurs at $T_s\approx 90$\,K, but, unlike other iron-based superconductors, no magnetic order occurs down to $T=0$, displaying peculiar electronic properties associated with the nematicity~\cite{McQueen09,Imai09,Baek15,Boehmer15,Watson15a,Wang16a,Wang16b,Hosoi16, Massat16}.  Significant enhancement of  $T_{\rm c}$ has been reported under hydrostatic pressure~\cite{Medvedev09,Sun16}, and more recently in the form of one-unit-cell-thick films on SrTiO$_3$~\cite{Wang12, He13}.  One of the most remarkable features of FeSe is its extremely small Fermi surface, which strongly deviates from predictions of first-principles calculations~\cite{Terashima2014,Watson2015}. The low-temperature Fermi surface of FeSe consists of a small hole pocket centered at the $\Gamma$ point and one or two small electron pockets at the $M$ point \cite{Terashima2014,Watson2015,ARPES2,ARPES3,ARPES4,ARPES5,ARPES6}. 
Strikingly, all these pockets are very shallow and the Fermi energies of the hole and electron pockets are extraordinarily small, with $\varepsilon_F^h\approx 15$\,meV and $\varepsilon_F^e\approx10$\,meV. %, i.e., comparable to $\Delta\sim 3$\,meV. 
It has been shown that FeSe is a multigap superconductor with two distinct superconducting gaps of $\Delta \sim 2.5$ and 3.5\,meV~\cite{Kasahara2014}. 
The comparable energy scale of $\varepsilon_{\rm F}$ and $\Delta$ indicates that superconductivity in FeSe is deep in the BCS-BEC crossover regime~\cite{Kasahara2014, ARPES1}. Recent observation of giant superconducting fluctuations  by far exceeding the standard Gaussian theory and a possible pseudogap formation well above $T_c$ have been attributed to distinct signatures of the crossover~\cite{Kasahara2016}.  

Furthermore, FeSe has two unique features which may provide new insights into fundamental aspects of the physics of BCS-BEC crossover. The first feature is the electronic structure: FeSe is a compensated semimetal with equal numbers of electron and hole carriers, and hence it is essentially a multiband superconductor. In fact, as discussed in Ref.~\cite{Chubukov}, this renders the crossover physics in FeSe distinguished from that in ultracold atomic gases.  However, although the superconducting gap structure has been reported to be very anisotropic, the detailed gap structure in each band is still unclear. The second feature concerns the fate of the superfluid when the spin populations are strongly imbalanced.  Although highly spin-imbalanced Fermi systems have been realized in ultracold atomic gases, the nature of the spin-imbalanced superfluid is still largely unexplored experimentally due to the difficulty in cooling the systems to sufficiently low temperature~\cite{Chevy10,Liao10,Gub12,Yosh07}. In superconductors, the spin imbalance is introduced through Zeeman splitting in an applied magnetic field. The magnitude of the spin imbalance $P=(N_{\uparrow}-N_{\downarrow})/(N_{\uparrow}+N_{\downarrow})$, where $N_{\uparrow}$ and $N_{\downarrow}$ are the numbers of up and down spins, respectively, is roughly estimated as $P\approx \mu_BH/\varepsilon_F$.  In almost all superconductors, $P$ is very small, $P\sim 10^{-3}$-$10^{-2}$ even near the upper critical field $H_{c2}$. On the other hand, in FeSe in the crossover regime, the Zeeman effect is particularly effective in shrinking the Fermi volume associated with the spin minority, giving rise to a highly spin-imbalanced phase where $\varepsilon_F\sim \Delta \sim\mu_BH$. Recent experiments of FeSe report a possible field-induced phase (dubbed $B$-phase) in the low-temperature and high-field region in  the $H$-$T$ phase diagram~\cite{Kasahara2014}, but very little is known about its nature.

To shed light on the above issues, the detailed knowledge of the quasiparticle excitations in the superconducting state is crucial. Here we measured the thermal conductivity $\kappa_{xx}$ and the thermal Hall conductivity $\kappa_{xy}$ up to $\mu_0H=20$\,T. % above $H_{c2}$.  %Both are unique transport quantities that do not vanish in the superconducting state.  Cooper pairs do not carry the entropy and therefore do not contribute to the thermal transport.  
Both quantities are sensitive probes of the delocalized low-energy quasiparticle excitations~\cite{Izawa01,MatsudaJPC}.  %Thermal Hall conductivity is the nondiagonal element of the thermal conductivity tensor in a perpendicular field. While $\kappa_{xy}$ is purely electronic and the direct consequence of a transverse quasiprticle current, while $\kappa_{xx}$ includes both electronic and phonon contributions.  
Our results reveal that a highly anisotropic small superconducting gap opens in the electron Fermi-surface pocket whereas a more isotropic and larger gap forms in the hole pocket.  We also find that the quasiparticle scattering rate is strongly modified on entering the $B$-phase. We discuss that this field-induced phase is likely to present a highly anomalous inhomogeneous superconducting state that has not been addressed before, rather than a conventional Fulde-Ferrel-Larkin-Ovchinnikov (FFLO) state~\cite{FFLO, LO}.

\section{Experiments}

High-quality single crystals of FeSe were grown by the chemical vapor transport method~\cite{Boehmer2013}. 
%The temperature at which $\rho_{xx}$ goes to zero is $\approx8.0$\,K in the present sample. The resistivity at the superconducting onset $\rho_{xx}(T_c^+)\approx24\, \mu \Omega{\rm cm}$ is roughly two times larger than that reported in Ref.\,\cite{Kasahara2014}, and the magnitude of the magnetoresistance $\Delta \rho_{xx}(H)/\rho_{xx}(0)$ is nearly half. % of that reported in Ref.\,\cite{Kasahara2014}. 
Both $\kappa_{xx}$ and $\kappa_{xy}$ were measured on the same crystal by the steady-state method, applying the thermal current \bm{$\dot{q}$} in the $ab$ plane with $\bm{\dot{q}}\parallel\bm{x}$ for $\bm{H}\parallel c$. The thermal gradients $-\nabla_xT \parallel\bm{x}$ and $-\nabla_yT \parallel\bm{y}$ were detected by RuO$_2$ thermometers, and $\kappa_{xx}=w_{xx}/(w_{xx}^2+w_{xy}^2)$ and $\kappa_{xy}=w_{xy}/(w_{xx}^2+w_{xy}^2)$ were obtained from the thermal resistivity $w_{xx}=\nabla_xT/q$ and thermal Hall resistivity $w_{xy}=\nabla_yT/q$. The effect of misalignment of the Hall contacts was eliminated by reversing the magnetic field at each temperature. We additionally determined electrical longitudinal and Hall resistivities, $\rho_{xx}$ and $\rho_{xy}$, respectively, in the same setup. The experimental setup is schematically presented in the inset of Fig.~1(b).

\section{Results}

\begin{figure}[t]
	\begin{center}
		\includegraphics[width=1.0\linewidth]{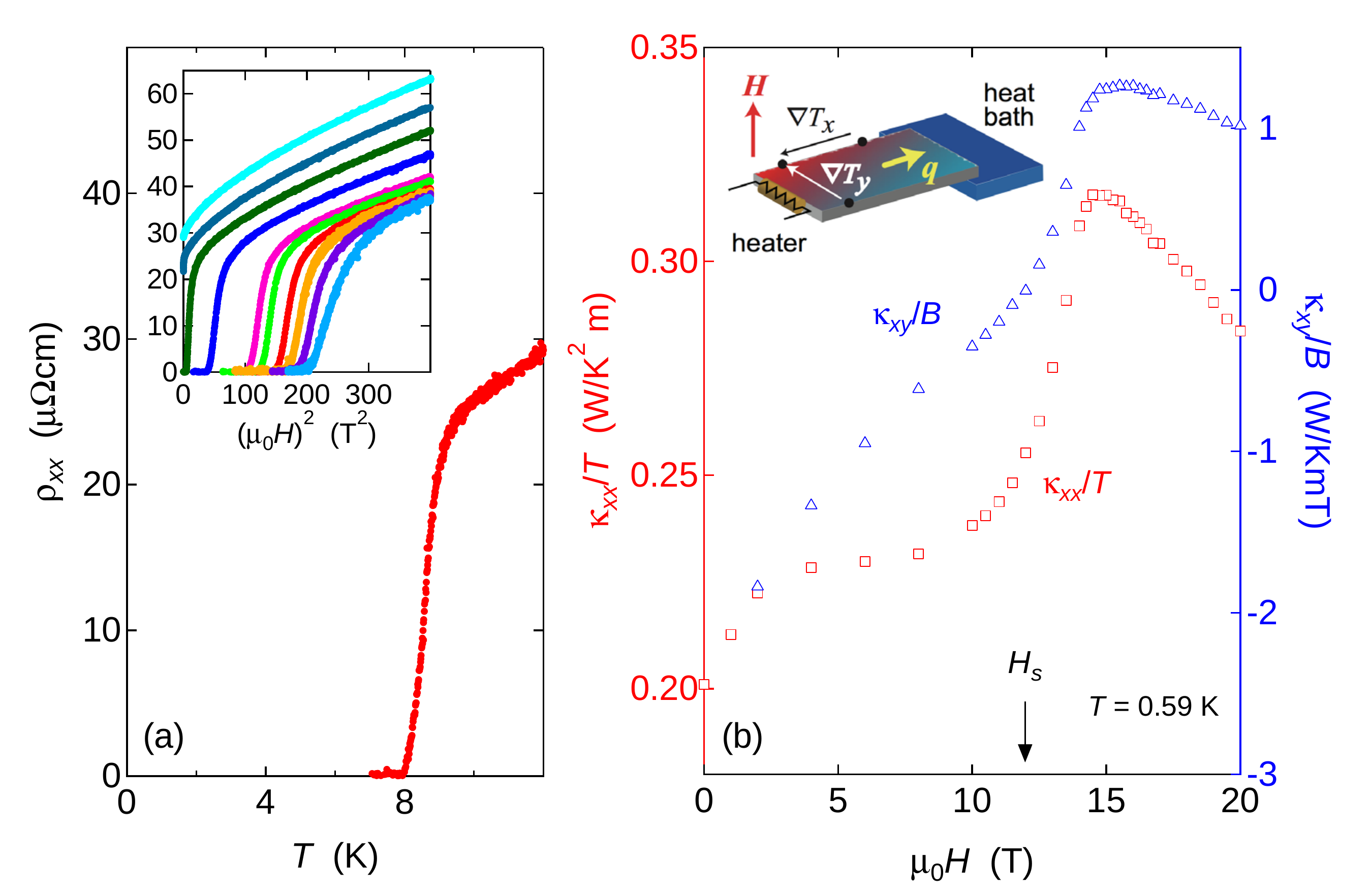}
	\end{center}
\caption{(Color online)
(a) Temperature dependence of the in-plane resistivity $\rho_{xx}(T)$ in zero field. The inset shows the field dependence of $\rho_{xx}(H)$ for $\bm{H}\parallel c$. From bottom to top, $T=0.31$, 0.61, 1.0, 1.4, 2.1, 2.6, 4.7, 6.6, 9.1, and 12\,K. (b)  Field dependence of $\kappa_{xx}/T$  (red squares, left axis) and  $\kappa_{xy}/B$ (blue triangles, right axis) at 0.59\,K. The inset shows a schematic of the measurement setup for the thermal-conductivity tensor. 	}
	\label{fig:Fig1}
\end{figure}

\subsection{Superconducting Gap Structure}
Figure\,1(a) and its inset show the $T$ and $H$ dependences of the in-plane resistivity $\rho_{xx}$, respectively. The temperature at which $\rho_{xx}$ goes to zero is $\approx8.0$\,K in the present sample. The resistivity at the superconducting onset $\rho_{xx}(T_c^+)\approx24\, \mu \Omega{\rm cm}$ is roughly two times larger than that reported in Ref.~\cite{Kasahara2014}, and the magnitude of the magnetoresistance $\Delta \rho_{xx}(H)/\rho_{xx}(0)$ is roughly halved. % of that reported in Ref.\,\cite{Kasahara2014}. 
Figure\,1(b) depicts the $H$ dependence of $\kappa_{xx}/T$ and $\kappa_{xy}/B$ (with $B=\mu_{0}H$) at $T=0.59$\,K.  
%Generally, 
Usually, $\kappa_{xy}$, which is the nondiagonal element of the thermal-conductivity tensor in a perpendicular field, is purely electronic.
%, %and the direct consequence of a transverse quasiparticle current while $\kappa_{xx}$ includes both electronic and phonon contributions. 
The ratio of $\kappa_{xy}/T$ and electrical Hall conductivity, $\sigma_{xy}=\rho_{xy}/(\rho_{xx}^2+\rho_{xy}^2)$, is $(\kappa_{xy}/T)/\sigma_{xy}=1.05L_0$ at $\mu_0H=20$\,T, where $L_0=\frac{\pi^2}{3}\left(\frac{k_B}{e}\right)$  is the Lorenz number, indicating that the Wiedemann-Franz law holds within experimental error. 
At low fields,  $\kappa_{xx}/T$ increases with convex curvature, compatible with the $\sqrt{H}$ dependence expected for a nodal gap (see below), and becomes nearly independent of $H$ between $\sim4$ and 10\,T. 
%$\kappa_{xy}/B$  increases nearly linearly at low fields and changes sign 
Above 2\,T, $\kappa_{xy}/B$ decreases in magnitude and then changes sign %around 12 T
from negative to positive around $\mu_0H_s\approx12\,$T.  Above $H_s$, both $\kappa_{xx}/T$ and $\kappa_{xy}/B$ increase steeply and decrease after exhibiting %broad maxima 
a narrow plateau-like ridge at around 15\,T.  %These broad maxima will be discussed later. 

The initial increase of $\kappa_{xx}/T$ with a convex curvature at low fields is a signature of superconductivity with line nodes, %signature of superconductors with line nodes, 
where the  Doppler shift of the quasiparticle spectrum in the presence of supercurrents around vortices gives rise to a  $\sqrt{H}$ increase in the population of delocalized quasiparticles \cite{MatsudaJPC}.  The superconducting gap structure of FeSe is subject of extense discussion. A clear signature of the line nodes in the superconducting gap has been reported for very clean single crystals~\cite{Kasahara2014} or thin films~\cite{Song2011} probed by scanning tunneling spectroscopy, thermal conductivity and magnetic penetration depth. In less clean crystals, however, a superconducting state with finite  energy gap  has been reported~\cite{Dong2009}.  Moreover, even in very clean crystals, the nodes have been reported to disappear near the twin boundaries \cite{Watashige2015}.  These results indicate that the line nodes are not symmetry protected and can be easily lifted by impurity and boundary scattering. %by microwave conductivity, thermal conductivity, specific heat, {\it etc}.  
%These suggest that the line nodes in FeSe are not symmetry protected but accidental ones. 
%In fact, recent STS study demonstrates that nodes are lifted by the presence of twin boundaries~\cite{Watashige15}
%In fact, it has been shown that line nodes in FeSe can easily be lifted by impurity scattering and the presence of twin boundaries \cite{Watashige2015}. 
%Therefore, the low-field convex curvature of $\kappa_{xx}/T$ in Fig.\,1b} may be due to the large gap anisotropy rather than line nodes \cite{Bourg2016}. 
The observed  low-field convex curvature of $\kappa_{xx}/T$ in Fig.\,1(b) indicates the presence of line nodes or large superconducting gap anisotropy.  Similar field dependence has been reported in FeSe with similar residual resistivity and magnetoresistance \cite{Bourg2016}. We note that this low-field behavior of $\kappa_{xx}/T$  differs from that reported in FeSe with much smaller residual resistivity, in which $\kappa_{xx}/T$ decreases rapidly with magnetic field~\cite{Kasahara2014}, possibly be because in very clean FeSe, the extremely long quasiparticle mean free path is reduced by vortex scattering at low fields~\cite{Kasahara2005}. 

The nearly step-like behavior in the field dependence of $\kappa_{xx}/T$ upon approaching 15\,T is attributed to the multiband nature of superconductivity in FeSe~\cite{Solo2002,Boaknin2003,Yamashita2009}. A substantial portion of quasiparticles, which can be easily excited over the smaller gap, is already present at fields well below $H_s$, and therefore, $\kappa_{xx}/T$ at high fields above $\sim4$\,T is governed by the larger gap. Note that a nearly $H$-independent behavior at low fields and a subsequent steep increase of $\kappa_{xx}/T$ with a concave curvature is observed in fully gapped superconductors, where quasiparticles are localized in the vortex core and thus are unable to transport heat until these vortices overlap each other~\cite{MatsudaJPC}. 

\subsection{Field-induced Superconducting Phase} 

\begin{figure*}[t]
	\begin{center}
		\includegraphics[width=0.95\linewidth]{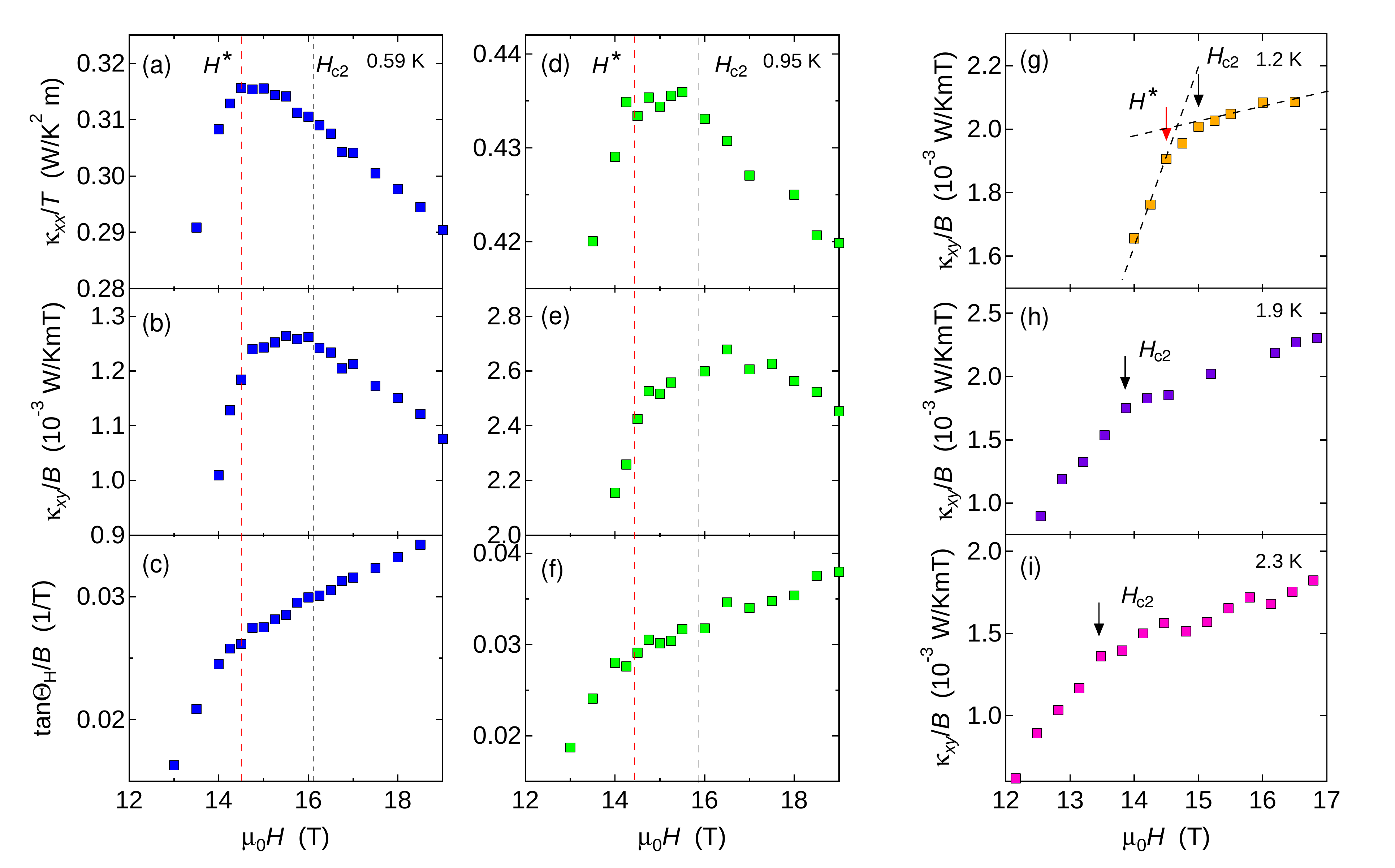}
	\end{center}
\caption{(Color online)
	%{Thermal conductivity, thermal Hall conductivity and thermal Hall angle near the upper critical field.}		
Field dependences of (a) $\kappa_{xx}(H)/T$, (b) $\kappa_{xy}/B$, and (c) $\tan\Theta_H/B\equiv\kappa_{xy}/(\kappa^{\rm qp}_{xx}B)$ at $T=0.59$\,K.  (d), (e) and (f) are the results at $T=0.95$\,K. (g), (h) and (i)  show $\kappa_{xy}/B$ at $T=1.2$, 1.9, and 2.3\,K, respectively.  
$\kappa_{xx}^{\rm qp}$ is obtained by subtracting $\kappa_{xx}^{\rm ph}$ from $\kappa_{xx}$, where $\kappa_{xx}^{\rm ph}$ is estimated by using the Wiedemann-Franz law in the normal state at 20\,T, $\kappa_{xx}^{\rm ph}/T=\kappa_{xx}(20\,{\rm T})/T-L_0\sigma_{xx}(20\,{\rm T})$, and is assumed to be field independent. The dashed red and black lines represent magnetic fields, %kink anomalies, 
which correspond to $H^*$ and $H_{c2}$, respectively. The dashed lines in (g) represent the linear extrapolations below $H^*$ and above $H_{c2}$.  Our detailed measurements of the thermal transport coefficients, $\kappa_{xx}$ and $\kappa_{xy}$, 
with small steps in the field scans allow us to observe two distinct anomalies at $H^\ast$ and $H_{c2}$, which have been missed in a previous study~\cite{Bourg2016}. 
In $\tan\Theta_H/B$, no discernible anomaly is observed at $H_{c2}$. 
The observed kink in $\kappa_{xx}(H)/T$ at $H^*$ is distinct from the previously reported observations in very clean FeSe \cite{Kasahara2014}, where $\kappa_{xx}/T$ exhibits a tiny peak at $H^*$. This may be caused by the very large magnetoresistance in the cleaner sample, which leads to the rapid decrease of $\kappa_{xx}/T$ at low $H$ and above $H_{c2}$, smearing out the rapid increase just below $H^*$ and features at $H_{c2}$.
}	\label{fig:Fig2}
\end{figure*}

Figures\,2(a), (b), and (c) depict $H$ dependences of $\kappa_{xx}/T$, $\kappa_{xy}/B$, and of the quasiparticle thermal Hall angle divided by $B$, $\tan\Theta_H/B\equiv\kappa_{xy}/(\kappa^{\rm qp}_{xx}B)$, respectively, at $T=0.59$\,K, while Figs.\,2(d), (e), and (f) show the results at $T=0.95$\,K. Two distinct anomalies (kinks) can be identified in both $\kappa_{xx}/T$ and $\kappa_{xy}/B$. We first point out that the field marked by the dashed black line corresponds to  $H_{c2}$ because of the following reasons. First, above this field, $\kappa_{xx}/T$ decreases rapidly with $H$, which is consistent with the large positive magnetoresistance in the normal state~\cite{Kasahara2007} [inset of Fig.\,1(a)].  In addition, the fact that $\kappa_{xx}/T$ is nearly independent of $H$ below this field definitely indicates that the system at lower fields is not in the normal state.  Second, the thermal conductivity has no  fluctuation correction, in contrast to the resistivity, magnetic susceptibility, and specific heat, which all are subject to fluctuations, and, hence, often exhibits a kink anomaly at the mean-field upper critical field \cite{Vish01,Okazaki08}.  Third, as will be shown later, this field is close to the superconducting onset temperature determined by the use of resistivity data. 
The fact that the admittedly weak kinks at $H^*$ and $H_{c2}$ occur at similar fields for the different measurements at temperatures 0.59 and 0.95 K lends strong confidence to the identification of $H^*$ and $H_{c2}$.

Figures\,2(g), (h), and (i) show the $H$ dependence of $\kappa_{xy}/B$ at higher temperatures.  At $T=1.2$\,K, no clear kink anomalies can be resolved, but $\kappa_{xy}/B$ deviates from the linear lines extrapolated from the lower and higher field regimes, as shown in Fig.\,2(g). At $T=1.9$ and 2.3\,K, only one kink anomaly is observed, which is attributed to $H_{c2}$ because, as will be shown later, these fields are  close to the onset of superconductivity.

\begin{figure}[t]
	\begin{center}
	%\hfill
		\includegraphics[width=0.9\linewidth]{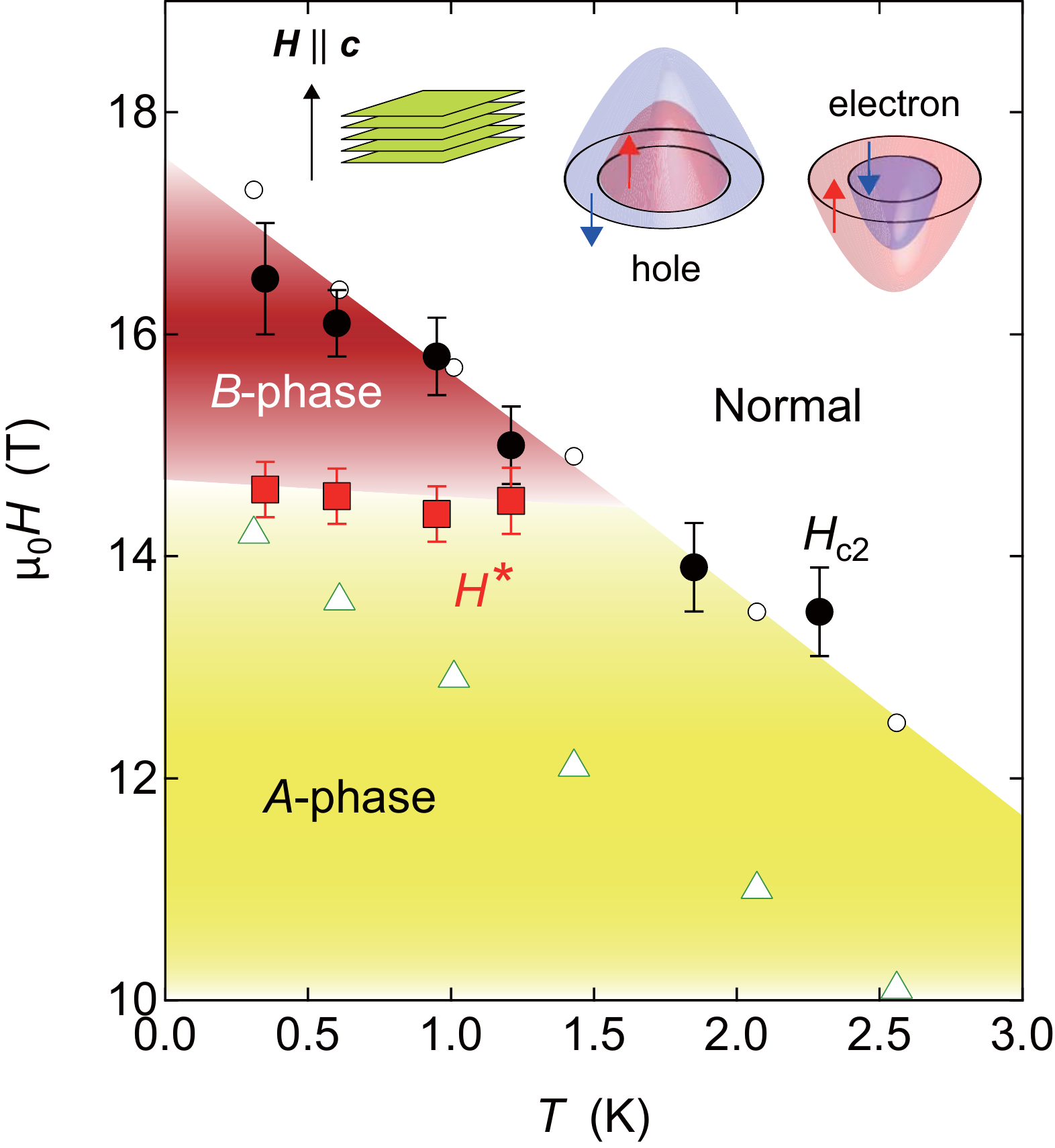}
	\caption{(Color online)	
	{Field-temperature ($H$-$T$) phase diagram of FeSe in perpendicular field determined by the thermal-transport measurements.} 
	Black filled circles represent the upper critical field $H_{c2}$ and filled red squares represent $H^*$ that separates $A$- and $B$-phases. The black open circles represent superconducting onset temperatures at which the resistivity becomes 90\% of the value extrapolated from the normal-state value under magnetic field, and the green open triangles represent the irreversibility fields determined by zero resistivity [inset of Fig.~1(a)].   
The inset illustrates the electronic structure of hole and electron pockets at fields around $H^*$. Both Fermi surfaces are highly spin imbalanced. 			
		}
		\label{fig:Fig3}
	\end{center}
\end{figure}

Figure\,3 displays the $H$-$T$ phase diagram. 
The magnitudes of $H^*$ and $H_{c2}$ in the present crystal are slightly larger than those reported in Ref.~\cite{Kasahara2014}.  
We note that the superconducting onset temperature (black open circles) is close to $H_{c2}$ determined by the thermal transport coefficients, (black closed circles) supporting that the higher-field anomaly shown by the dashed black line in Figs.\,2(a)-(i) is indeed due to $H_{c2}$. In contrast to $H_{c2}$, $H^*$ is nearly temperature independent and disappears when it meets the $H_{c2}$ line. 
We stress that this fact is an important additional feature establishing the $H$-$T$ phase diagram, which was not seen in our previous work~\cite{Kasahara2014}.
We further note that the $B$-phase above $H^*$ is not related to magnetic ordering, because of the following reasons. First, magnetic-torque measurements show no anomaly at $H^*$ \cite{Kasahara2014}.  Second, the size of the hole pocket determined by magnetic quantum oscillations above 20\,T coincides with that determined from angle-resolved photoemission in zero field, suggesting the absence of band folding \cite{Watson15}. Third, the $H^*$ line is observable only in the superconducting regime.  

It has been reported that $\rho_{xy}$ is positive and increases linearly with $H$ and that the magnetoresistance  increases with $H$ as $\Delta\rho_{xx}(H)/\rho_{xx}(0)\propto H^2$ at $\mu_0H\gtrsim18$\,T \cite{Watson2015}. Since FeSe is a compensated metal, with the same number of hole and electron charge carriers and only a single hole pocket,  these $H$ dependences indicate that the transport properties in the high-field region are governed by that hole band \cite{Pippard,Kasahara2007}. In this situation, the scattering time $\tau_{\rm qp}$ of the quasiparticles is proportional to the thermal Hall angle, $\tan\Theta_H/B\propto\tau_{\rm qp}$. The clear kink at $H^*$ in $\tan\Theta_H/B$ [Figs.\,2(c) and (f)] indicates that the quasiparticle scattering mechanism changes when crossing the $H^*$ line, suggesting a modification of the superconducting gap structure. At $H_{c2}$, $\tan\Theta_H/B$ shows no discernible anomaly, while $\kappa_{xx}/T$ shows a clear kink as shown in Figs.\,2(a) and (d). Since $\kappa_{xx}$ is proportional to the product of $\tau_{\rm qp}$ and the quasiparticle density $n_{\rm qp}$, $\kappa_{xx} \propto n_{\rm qp}\tau_{\rm qp}$, these results show that $\tau_{\rm qp}$ barely changes while  $n_{\rm qp}$ is reduced below $H_{c2}$.   Thus, the observed anomalies in the thermal transport coefficients  strongly suggest the emergence of a new high-field superconducting phase ($B$-phase) which is distinctly different from the $A$-phase at lower fields.

\section{Discussion}

Although multigap superconductivity and a very anisotropic or nodal gap structure in FeSe have been suggested by several experiments, the detailed gap structure of electron and hole pocket has remained unclear. The combined results of the field dependence of $\kappa_{xx}$ and the  sign of $\kappa_{xy}$ provide important information on this issue.
The sign change of $\kappa_{xy}$ at $H_s$ evidences that the quasiparticles responsible for the thermal conduction are electron-like at low fields and  hole-like at high fields. Therefore, the large gap anisotropy found at low fields should correspond to the electron pocket, whereas the concave increase above $H_s$ and saturation of $\kappa_{xx}/T$ below $H_s$ indicate that the superconducting gap of the hole pocket is more isotropic and that its amplitude is larger than that of the electron pocket. %~\cite{Kasahara2014,Watashige2015}. 

Next we discuss the field-induced superconducting $B$-phase.  We stress that this phase represents a highly unusual superconducting state because both electron and hole pockets are largely spin imbalanced with $P\approx \mu_BH^*/\varepsilon_F\approx 0.1$, as illustrated in the inset of Fig.\,3. 
The nearly temperature-independent behavior of $H^*$ suggests that this anomaly is not an artifact of a single broad peak of the $H_{c2}$ anomaly, because in that case the two anomalies at $H^*$ and $H_{c2}$ should follow parallel lines.
One of the candidates of the $B$-phase may be a Fulde-Ferrell-Larkin-Ovchinnikov (FFLO) phase~\cite{FFLO,LO}, where the Cooper-pair formation occurs between Zeeman-split parts of the Fermi surface and a new pairing state $({\bm k} \uparrow, -{\bm k}+{\bm q}\downarrow)$ with finite ${\bm q}$ is realized. The FFLO state has been reported in layered heavy-fermion and organic superconductors for magnetic fields parallel to the 2D plane \cite{Shimahara, Wosnitza}. However, we want to point out that the $B$-phase is unlikely to be a conventional FFLO phase.  In fact, the $B$-phase coexists with the vortex-liquid state where the resistivity is finite. This is different from heavy-fermion and organic systems, where possible FFLO phases occur within the vortex-lattice state. Moreover, the FFLO state is stabilized when a large part of the spin-up Fermi surface is connected to the spin-down surface by a single ${\bm q}$ vector.  For the lowest Landau-level solution, the superconducting order parameter changes spatially as $\Delta(\bm{r})\propto \cos({\bm q}\cdot {\bm r})$ where ${\bm q}\parallel {\bm H}$. However, since  the Fermi surface of layered FeSe is cylindrical, the spin-up Fermi surface cannot be connected to a large part of the spin-down Fermi surface by a single ${\bm q}$  for $\bm{H}\parallel c$. This analysis leads us to consider that the $B$-phase may be an FFLO phase involving  higher Landau levels, where ${\bm q}$ lies in the two-dimensional plane.  Moreover, what makes FeSe distinctly different from ultracold atomic gases is  that FeSe is  a multiband system where the interplay between the formation of bound pairs and superconductivity may lead to rich physics \cite{Chubukov}.  Thus, the field-induced $B$-phase appears to be a previously unidentified kind of inhomogeneous superconducting state with an unprecedented large spin imbalance, which apparently is associated with the BCS-BEC crossover in a multiband system.

\section{Summary}
In summary, we measured thermal longitudinal and thermal Hall conductivities in FeSe %, which is a unique and fascinating platform to study a novel superconducting state 
in the BCS-BEC crossover regime. The field dependences of thermal transport coefficients imply that the quasiparticle structure is strongly band-dependent; a highly anisotropic small superconducting gap forms at the electron Fermi-surface pocket whereas a more isotropic and larger gap forms at the hole pocket. The analysis of the thermal Hall angle provides evidence for the emergence of a highly anomalous inhomogeneous superconducting state, which is likely to be caused by the extremely high spin polarization. The strongly band-dependent quasiparticle structure and the emergent highly spin polarized phase may be important and unique features of the BCS-BEC crossover in the multiband superconducting system. 
%Thus, FeSe is a unique and fascinating platform to study a novel superconducting state in the BCS-BEC crossover regime. 

%Therefore FeSe offers a unique and fascinating platform to study the crossover physics in the multiband  systems.  
%Here, to obtain information on the quasiparticle excitations in the superconducting state of FeSe, 

%\begin{Acknowledgment}

%\acknowledgment
\section*{Acknowledgements}

We thank Y. Yanase, R. Ikeda, S. Fujomoto for discussions. 
This work was supported by Grants-in-Aid for Scientific Research (KAKENHI) (No. 25220710, No. 15H02106, No. 15H03688), Grants-in-Aid for Scientific Research on Innovative Areas ``Topological Materials Science" (No. 15H05852) from Japan Society for the Promotion of Science (JSPS), and HLD at HZDR, a member of the European Magnetic Field Laboratory (EMFL). 

%\end{Acknowledgment}

\end{document}